\documentclass{PoS}

\title{Son of X--Shooter: a multi--band instrument for a multi--band universe}

\ShortTitle{SOXS}

\author{\speaker{R. Claudi}\thanks{riccardo.claudi@inaf.it}\ $^a$, S. Campana$^b$, P. Schipani$^c$, M. Aliverti$^b$, A. Baruffolo$^a$, S. Ben-Ami$^{d,q}$, F. Biondi$^a$, A. Brucalassi$^{e,k}$, G. Capasso$^c$, R. Cosentino$^{f,i}$, F. D'Alessio$^g$, P. D'Avanzo$^b$, O. Hershko$^d$, H. Kuncarayakti$^{h,p}$, M. Munari$^i$, A. Rubin$^d$, S. Scuderi$^i$, F. Vitali$^g$, J. Achr\'en$^j$, J. Antonio Araiza-Duran$^k$, I. Arcavi$^l$, A. Bianco$^b$, E. Cappellaro$^a$, M. Colapietro$^c$, M. Della Valle$^c$, O. Diner$^d$, S. D'Orsi$^c$, D. Fantinel$^a$, J. Fynbo$^m$, A. Gal-Yam$^d$, M. Genoni$^b$, M. Hirvonen$^n$, J. Kotilainen$^{h,o}$, T. Kumar$^o$, M. Landoni$^b$, J. Lehti$^n$, G. Li Causi$^g$, L. Marafatto$^a$, S. Mattila$^o$, G. Pariani$^b$, G. Pignata$^k$, M. Rappaport$^d$, M. Riva$^b$, D. Ricci$^a$, B. Salasnich$^a$, S. Smartt$^p$, M. Turatto$^a$, R. Zanmar Sanchez$^i$, H.U. K\"aufl$^e$, M. Accardo$^e$\\
        $^a$INAF-- Osservatorio Astronomico di Padova, vicolo Osservatorio, 35122 Padova, Italy\\
        $^b$INAF --Osservatorio Astronomico di Brera, Via Bianchi 46, I-23807 Merate (LC), Italy\\
        $^c$ INAF -- Osservatorio Astronomico di Capodimonte, Salita Moiariello 16, I-80131, Napoli, Italy\\
        $^d$ Weizmann Institute of Science, Herzl St 234, Rehovot, 7610001, Israel\\
        $^e$ ESO, Karl Schwarzschild Strasse 2, D-85748, Garching bei M\"unchen, Germany\\
        $^f$ FGG-INAF, TNG, Rambla J.A. Fern\'andez P\'erez 7, E-38712 Bre$\tilde{n}$a Baja (TF), Spain\\
        $^g$ INAF -- Osservatorio Astronomico di Roma, Via Frascati 33, I-00078 Monte Porzio Catone, Italy\\
        $^h$ Finnish Centre for Astronomy with ESO (FINCA), FI-20014 University of Turku, Finland\\
        $^i$ INAF -- Osservatorio Astronomico di Catania, Via S. Sofia 78 30, I-95123 Catania, Italy\\
        $^j$ Incident Angle Oy, Capsiankatu 4 A 29, FI-20320 Turku, Finland\\
        $^k$ Universidad Andres Bello, Avda. Republica 252, Santiago, Chile\\
        $^l$ Tel Aviv University, Department of Astrophysics, 69978 Tel Aviv, Israel\\
        $^m$ DARK Cosmology Center, Juliane Maries Vej 30, DK-2100 Copenhagen, Denmark\\
        $^n$ ASRO (Aboa Space Research Oy), Tierankatu 4B, FI-20520 Turku, Finland\\
        $^o$ Tuorla Observatory, Department of Physics and Astronomy, FI-20014 University of Turku, Finland\\
        $^p$ Astrophysics Research Centre, Queen's University Belfast, Belfast, County Antrim, BT7 1NN, UK\\
        $^q$ Harvard Smithsonian Center for Astrophysics, Cambridge, USA
        }

\abstract{Son Of X-Shooter (SOXS) will be a new instrument designed to be mounted at the Nasmyth--A focus of the ESO 3.5 m New Technology Telescope in La Silla site (Chile). SOXS is composed of two high-efficiency spectrographs with a resolution slit product  4500, working
in the visible (350 -- 850 nm) and NIR (800 -- 2000 nm) range respectively, and a light imager in the visible (the acquisition camera usable also for scientific purposes).  The science case is very broad, it ranges from moving minor bodies in the solar system, to bursting young stellar objects, cataclysmic variables and X-ray binary transients in our Galaxy, supernovae and tidal disruption events in the local Universe, up to gamma-ray bursts in the very distant and young Universe, basically encompassing all distance scales and astronomy branches.  At the moment, the instrument passed the Preliminary Design Review by ESO (July 2017) and the Final Design (with FDR in July 2018).}

\FullConference{Frontier Research in Astrophysics - III (FRAPWS2018)\\
		28 May - 2 June 2018\\
		Mondello (Palermo), Italy}

\begin{document}

\section{Introduction}
The research on transients has expanded significantly in the past two decades, leading to some of the most recognized discoveries in astrophysics (e.g. gravitational wave events, gamma-ray bursts, super-luminous supernovae, accelerating universe). Nevertheless, so far most of the transient discoveries still lack an adequate spectroscopic follow-up. Thus, it is generally acknowledged that with the availability of so many transient imaging surveys in the next future, the scientific bottleneck is the spectroscopic follow-up observation of transients. Within this context, SOXS aims to significantly contribute bridging this gap. It will be one of the few spectrographs on a dedicated telescope with a significant amount of observing time to characterize astrophysical transients. It is based on the concept of X-Shooter \cite{vernetetal2011} at the VLT but, unlike its ''father'', the SOXS science case is heavily focused on transient events. Foremost, it will contribute to the classifications of transients, i.e. supernovae, electromagnetic counterparts of gravitational wave events, neutrino events, tidal disruptions of stars in the gravitational field of supermassive black holes, gamma-ray bursts and fast radio bursts, X-ray binaries and novae, magnetars, but also asteroids and comets, activity in young stellar objects, and blazars and AGN.

In 2015 ESO selected SOXS out of 19 proposals in response to the ''Call for Scientific Projects for the NTT on the La Silla Observatory'', which had been issued in February 2014. SOXS has a key role in the new ESO strategy for the La Silla Observatory described in the ESO long term plan \cite{dezeeuw2016}, that envisages the dedication of the two telescopes operated by ESO in La Silla to specific topics. They are the study of the transient sky with SOXS at the NTT and the radial velocity studies for exoplanets with HARPS  \cite{mayoretal2003} and the new instrument NIRPS \cite{wildietal2017} at the 3.6m telescope.
As a sort of pathfinder to the SOXS science, a large fraction of the NTT observing time over the past few years has been dedicated to a public spectroscopic survey (the Public ESO Spectroscopic Survey of Transient Objects -- PESSTO \cite{smarttetal2015}) with 150 nights per year and a Large Programme (ePESSTO) with 200 nights over two years.
SOXS has very clear synergies with many other existing or upcoming major facilities. Many ground- or space-based facilities for searching new transients are operating, starting or are planned in the near future. Among them are GAIA \cite{gaiacollaborationetal2016}, PanSTARRS \cite{kaiseretal2010}, Zwicky Transient Factory \cite{bellm2018}, LSST \cite{ivezicetal2008} and EUCLID \cite{laureijsetal2012} for optical searches, Swift \cite{gehrelsetal2004}, Fermi \cite{atwoodetal2009}, SVOM\footnote{http://www.svom.fr/en/\#filter=.accueil}, MAGIC \cite{lorenz2004} and CTA\footnote{https://www.cta-observatory.org} for high-energy objects and, in the newly emerged field of non electromagnetic messengers, LIGO/VIRGO \cite{theligocollaborationetal2011} for gravitational waves and KM3NET\footnote{http://www.km3net.org} and ICECUBE\footnote{https://icecube.wisc.edu} for neutrinos. 
SOXS will simultaneously cover the electromagnetic spectrum from 0.35 to $2.0 \mu$m using two arms (UV--VIS and NIR) with a product slit-resolution of $\sim 4500$. The throughput will enable to reach a S/N$\sim 10$ in a 1-hour exposure of an R$=20$ mag point source. 

SOXS is supposed to start the operations in 2021.
The SOXS consortium is in charge of the realization of the instrument, with duties extending also over the next operation phase, within the framework of an agreement with ESO. The consortium is supposed to provide the user support through a helpdesk. This includes providing tools for the community (e.g. the ETC, tools to construct and submit the observing blocks). The consortium will further provide scheduling and merging of targets from the GTO and the regular ESO programmes. The transient programmes will be supported through dynamical scheduling and real-time interaction with the telescope operators. Data reduction to the 1D extracted spectrum will be done by the SOXS consortium and delivered to the ESO archive. The consortium will provide a public pipeline for the SOXS data.
In return of the efforts and investments, the SOXS consortium will be remunerated with 900 NTT nights over 5 years. The ESO community will access the rest of the NTT observing time. The consortium will be granted a proprietary period for their data. ESO will provide telescope operators and day-time maintenance and support on site.
The SOXS consortium structure has evolved since the proposal, including new partners beside the PI Institute Istituto Nazionale di AstroFisica (INAF). They are: Department of Particle Physics and Astrophysics, Weizmann Institute of Science (Israel); University Andres Bello and Millennium Institute for Astrophysics (Chile); FINCA - Finnish Centre for Astronomy with ESO \& Turku University (Finland); Queen's University Belfast (Ireland); Tel Aviv University (Israel); Niels Bohr University (Denmark). 

\section{The Science Case}
\label{sec:sciencecase}
\begin{figure}
 \includegraphics[width=.5\textwidth]{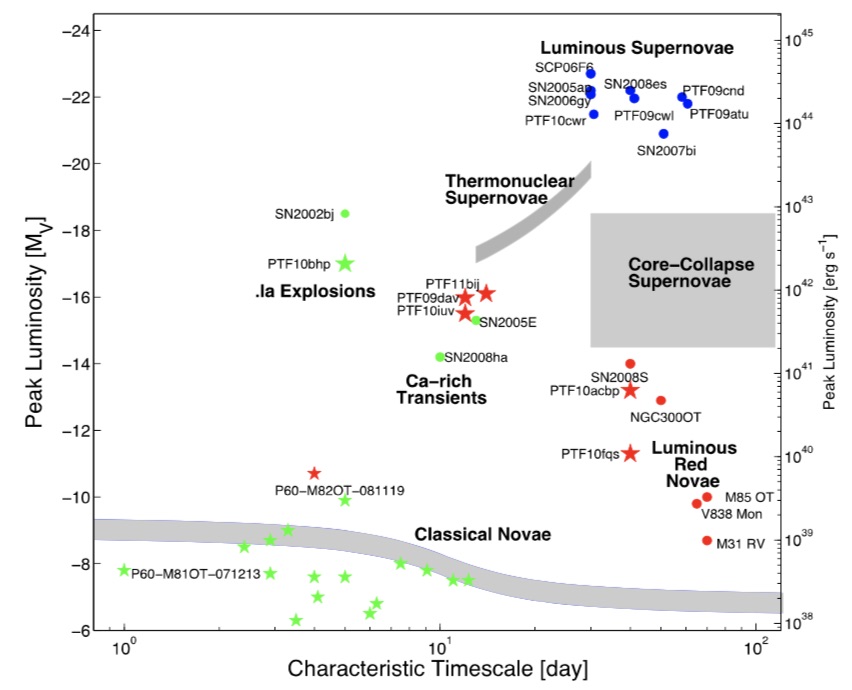}
 \includegraphics[width=.5\textwidth]{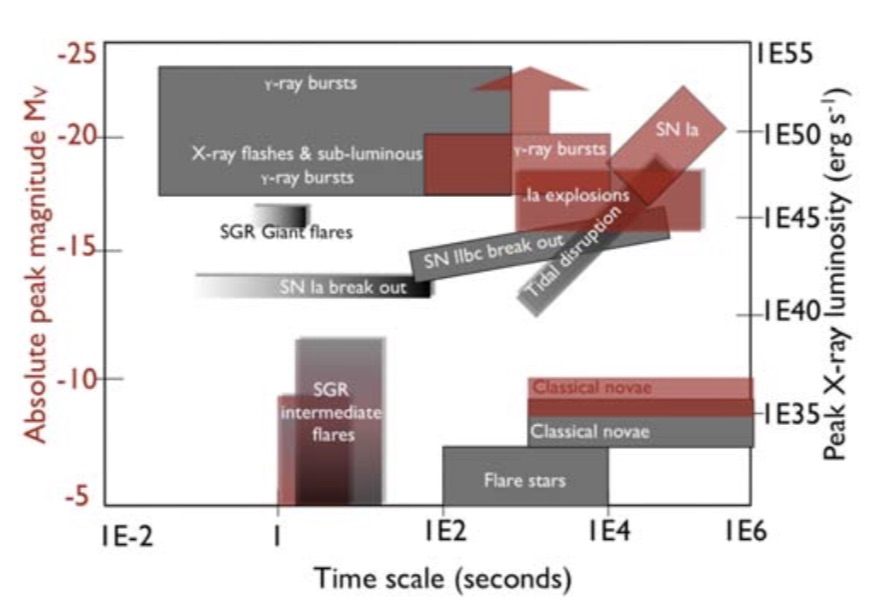} 
\caption{{\it Left}: Peak luminosity as a function of characteristic decay/variability timescale for luminous optical transients and variables \cite{kulkarni2012}. {\it Right}: The absolute optical magnitude (left labels and red colours) or X-ray luminosity (right labels and in grey) is plotted vs. the variability time \cite{jonkeretal2012})}
\label{lab:fig1}
 \end{figure}

Variable and transient objects encompass all astronomical distances ranging from comets and asteroids, exo-planets, stars (normal and compact), novae, supernovae, blazars or tidal disruption events. Up to now a lot of transient surveys are active in a very broad wavelength range from high energy ranges up to the infrared and most of these facilities will be active up to 2020 and beyond, and surely new will come.  
The discovery space in this research field is potentially immense, including virtually all astronomy disciplines (Figure \ref{lab:fig1}).

Much of the science extracted from these new windows will require the identification of electromagnetic counterparts, which would yield the redshift (and in turn the luminosity) of the object. The result is that the number of detected transients is very large and only a minimal fraction of them can be followed-up spectroscopically and thus characterised. 
Transients discovered up to now by current surveys have magnitudes brighter R $\lesssim \ 20$, that are well suited to be followed--up by a 4 m--class telescope like e.g. NTT where SOXS will be mounted. The main scientific drivers for such an instrument could be summarized in the following cases:

\noindent
{\bf Minor planets and asteroids}: Near--Earth objects (NEOs), Potentially Hazardous Asteroids (PHAs), Comets (long and short periods) and new comets and Centaurs, Trojans (Jupiter and Neptune's) and the brightest Trans-Neptunian Objects are all remnants of the outer proto--planetary disk. 
The physical characterization of these objects with the optical/NIR spectroscopy is fundamental to understand of the conditions existing in the early Solar System and of the accretion processes which governed the planetary formation at large heliocentric distances. 
Optical and NIR spectra are essential to derive a taxonomic classification of the objects that can lead to an acceptable prediction of the albedo.
 The knowledge of the albedo is critical to determine the size of the objects, being given their optical brightness. 

\noindent
{\bf Exoplanet}: Since the discovery of the first exoplanet orbiting a solar star (51\ Peg\ b \cite{mayorandqueloz1995}) much effort has gone into the studies of atmospheres around these planets. The reason for this is that the atmospheres hide information of not only about the evolution and characterization of these planets, but also about the possibility for these planets to host life. The characterization of atmospheres of transiting exoplanets can be done in three ways: transmission spectroscopy, secondary eclipses and phase light curves. Among them, the transmission spectroscopy method \cite{brown2001} offers a unique opportunity to estimate the key constituents in the atmosphere. 
Of about 1,000 transiting exoplanets, only a small fraction are bright enough (V$<12$) to be suitable for atmospheric studies, but with TESS already operative and after the launch of Cheops (2019) and PLATO (2028) many more such systems around bright star will be found. The molecules that we will mainly be targeting with the SOXS are O$_2$, H$_2$O, CO$_2$ and CH$_4$. 
In addition, during secondary eclipse, the planet passes behind the star and we can measure the planet's radiation directly by subtracting the photometric measurement during an eclipse from the measurement before or after the eclipse.

\noindent
{\bf Young Stellar Objects}: Young stellar objects (YSOs) of low--to--intermediate mass accumulate most of their final mass during the so--called main accretion phase ($10^5$\ yr). After this stage, the star appears as a pre--main sequence object accreting matter from a circumstellar disk. Observations show that the disk accretion process takes place through rapid and intermittent outbursts (up to 4--5 mag), usually detected at optical/NIR wavelengths, which can be related to a sudden increase of the mass accretion rate by orders of magnitude. 
The mechanism responsible for the onset of the accretion outbursts is not known. 
The very uncertain picture of accretion outbursts stems not only from the small number of known eruptive variables (a few tens), but especially from the lack of a proper (multi--wavelength) spectral monitoring. SOXS will have the potential to tackle both these problems, by observing the eruptive objects whenever an outburst is detected, in a spectral range (optical/NIR) that provides simultaneous information on the central object, accretion flows, disk, and on the matter ejected in winds/jets. 

\noindent
{\bf Novae and Cataclysmic Variables}: 
Nova outbursts are caused by thermonuclear explosions on white dwarfs (WDs) in cataclysmic binary systems \cite{bodeandevans2008} producing an increases of the optical brightness of the WD by a factor $10^4 - 10^5$ within a few hours/weeks. This leads to the ejection of mass at high velocities (1,000-5,000 km s$^{-1}$). A fraction of the accreted material remains in steady hydrogen burning on the surface of the WD, powering a supersoft X-ray source (SSS) visible once the ejected envelope becomes transparent \cite{henzeetal2013}. 
The study of Novae is important because they are considered as possible progenitors of SNe Ia and  they have also an important role in the chemical evolution of the Milky Way \cite{paynegaposchkin1957, livioandtruran1994}.
Finally we note that Fermi/LAT survey has detected novae $\gamma$-ray emissions at energy $\geqslant 100$\ MeV \cite{abdoetal2010}, which were totally unexpected and are yet unexplained. 
Summarizing, SOXS will help to address the following points: {\it i)} constrain and model the filling factor and ejecta masses, {\it ii)} analyse the evolution and determine the origin of the narrow metal absorptions, {\it iii)} characterise across multiple wave-bands the various transition phases of novae, we need for medium/high resolution spectroscopy monitored with SOXS and in coordination with observations from space (e.g. Swift, Chandra).

\noindent
{\bf Intermediate Luminosity Transients}: Intermediate Luminosity Optical Transients (ILOTs; \cite{sokerandkashi2012}) have absolute magnitudes intermediate between classical novae and the most common SN types (R between $\sim10$ and $\sim15$). ILOTs can also be outbursts of massive stars, including super--AGBs, Luminous Blue Variables (LBVs) and Wolf--Rayet stars, and the so--called ''luminous red novae'' (LRNe). 
ILOTs are typically characterised by narrow--lined spectra, fast SED evolution and, unusually, wide emission in the NIR domain. The wide range of evolutionary scenarios and the small number of events identified for each class leave many open questions concerning the causes for instabilities, the effect of mass loss, the physics of eruption/explosion. For most of the ILOT the ejecta expansion velocity are moderate, from a hundred to a few thousands km s$^{-1}$, and therefore the good resolution of SOXS will be instrumental to recover the kinematics of the ejecta and their evolution in time, the pre--supernova mass loss history as well as to provide a definitive classification.

\noindent
{\bf Supernovae (Ia, CC)}:
Currently the tightest constraints on the Einstein's cosmological constant come from combining observations of Type Ia supernovae (SNe Ia) with observations from other probes, such as the Cosmic Microwave Background and galaxy clustering. 
The best way to address this issue is done with high S/N optical and near-infrared (NIR) observations of nearby SNe Ia, which are well within the reach of the NTT + SOXS.
Core-collapse SNe show a wide variety of observed properties that have been (partly) understood and related to the diversity in the physical parameters of the progenitor stars (below $40 - 50$\ M$_\odot$ stars \cite{hegeretal2003}). Key information for the modelling of core collapse SNe and to understand the realm of progenitors are:
\begin{itemize}
\item  the evolution of the spectral energy distribution (SED) that give information on the ejecta temperature and bolometric correction. 
 Regular optical/IR monitoring of core collapse SNe with SOXS will allow a major step forward for a self-consistent modelling of SNe.
\item the detection in the near IR, in particular in the H-band, of CoII lines that allow measuring the amount and location of radioactive material in the ejecta \cite{stritzingeretal2015}. 
An instrument like SOXS can allow us to obtain proper statistics for this kind of data.
\end{itemize}

\noindent
{\bf Gamma--ray bursts}: The Gamma Ray Burst (GRB) phenomenon itself is not fully understood, i.e. neither the precise progenitors systems nor the emission mechanism for the prompt and afterglow emission. Concerning the progenitor systems most is known about long GRBs (duration of the high energy transient $>2$\ s), where the association with massive stellar death seems solid. Nevertheless, there have been puzzling cases of long GRBs that were not associated with bright SNe \cite{fynboetal2006, dellavalleetal2006}. This implies that either massive stars can die without producing bright SNe or there are other progenitors for long GRBs than core-collapse of massive stars. An additional important uncertainty on long GRB progenitors is the role of metallicity. 
For this science the crucial capability is resolution (to resolve the sky-lines), efficiency (to record the faint and transient emission), stable flux calibration (to be able to extract the relevant astrophysical information, in particular the shape of the red damping wing), and, more importantly, rapidity since the afterglow flux steadily decays.

\noindent
{\bf Radio sky Transients and Fast radio Bursts}: Fast radio bursts (FRBs) represent one of the most tantalising recent discoveries \cite{lorimeretal2007, thorntornetal2013} in the study of the transient sky. They appear as non--repeating bright radio pulses of millisecond duration, having high flux density, an electron column density (called dispersion measures DM) well in excess of the expected Galactic contribution along the line of sight and an inferred very high occurrence rate, of order $\sim1$ thousand events/day on the whole sky. These features suggest that FRBs could arise from energetic events (isotropic emission of $10^{38}-10^{40}$\ erg), occurring at co--moving distances around 1--3 Gpc and having brightness temperature of order $10^{33}-10^{36}$\ K. Although Giant Flares from magnetars seem a favoured hypothesis in literature (see e.g. the review of Ref.\ \cite{kulkarnietal2014}), the nature of the sources associated to these events still remains largely unconstrained.
In this context a spectroscopic follow up of the FRBs in the optical/NIR bands appears exceptionally promising. Moreover, the prompt reaction capabilities to ToO events (like FRBs are), as well as the wide band and good spectral resolution are very effective ways for following up the few hundreds FRB events yr$^{-1}$ which are expected to be detected by the experiments running around 2020 at radio telescopes like LOFAR, Long Wavelength Array (LWA), Murchison Widefield Array (MWA), Australian Square Kilometer Array Pathfinder (ASKAP), and SKA1.

\noindent
{\bf Gravitational Waves and neutrino electromagnetic counterparts}: The identification of GW and neutrinos sources in the electromagnetic sky is providing a scientific breakthrough in several field of astrophysics \cite{abbotetal2017, pianetal2017, smarttetal2017}, as for example, in understanding the physical process involving the matter under extreme conditions or the birth and evolution of compact objects. Moreover, the detection of the short GRB electromagnetic (EM) signals as counterpart of a coincident GW signal provided strong constraints on the nature of these events. The detection of neutrinos and GWs related to an EM observation of SN represents an extraordinary probe to investigate the physics in the very interior of the exploding stars and the dynamics of the explosion. Furthermore, the coincident detection of a GW signal from the so--called kilonova radiation \cite{piranetal2013} have provided and will provide for future events information of the merger processes, as the light curves depend on the mass, velocity and geometry of the ejecta, while their opacity and spectral feature probe the ejecta composition \cite{barnesandkasen2013}.
Nowadays, iPTF\footnote{https://www.ptf.caltech.edu/iptf} has demonstrated that coupling imaging and spectroscopic observations allows us finding GRBs with error box of the order of 100 deg$^2$ \cite{singeretal2013, singeretal2015}. 

\section{The Science Requirements}
\label{sec:sciencereq}
The science requirements (SR) follow from the science case for SOXS described in Section\ \ref{sec:sciencecase}. They are listed here while the top level requirements that have been derived from them are given in the next sections.
\begin{description}
\item[SR01: Wavelength range] The wavelength range that is covered by the SOXS in a single shot shall cover at least from U to H bands (350-1750 nm, 320-2050 nm goal).
\item[SR02: Detector Quantum Efficiency]  The instrument shall be highly efficient to be able to obtain quantitative spectroscopic information on the faintest, point--like targets reachable at a 3.5\ m telescope at a resolution where the spectrum is just sky limited in about one hour of exposure time in the regions free of emission lines. The absolute efficiency shall be the one given by state of the art coatings, detectors and transmitting materials.
\item[SR03: Spectral resolution] It shall be intermediate to maximise sensitivity but still be high enough to allow for quantitative work on narrow emission and absorption lines under median La Silla seeing and to be relatively unaffected by atmospheric OH emission lines. Sky background, detector noise and OH line density considerations require that the resolution slit product is $>3500$ to permit accurate sky subtraction.
\item[SR04: Spectral format] The spectral format is cross-dispersed (or quasi), dictated by the requirements on spectral resolution and wavelength range. Efficiency considerations strongly favour prism cross--dispersion.
\item[SR05: Slit length] It shall be adequate to perform sky subtraction with sufficient accuracy within the constraints of the spectral format. A minimal length of 10 arcsec is envisaged.
\item[SR06: ADC] In view of its wide band-pass, the instrument shall be equipped with an ADC (Atmospheric Dispersion Compensator) for at least the UV-VIS spectral ranges for good spectrophotometric accuracy and the flexibility to orient the slit as required by the observing program.
\item[SR07: Observing efficiency] The instrument shall be designed for rapid acquisition and low observing and calibration overheads. There shall be no need to perform calibration during the night. \item[SR08: Calibration system] SOXS shall be equipped with its own calibration system to perform scientific and technical calibrations needed to remove the instrument signature and to maintain the instrument.
\end{description}

\section{The Top Level Requirements}
\label{sec:requirements}
Following the science case and the Science requirement a score of Top Level Requirements (TLR) are also defined in order to have a well defined path to be followed to fulfil the expectations of the transients community. 
\begin{description}
\item[TLR\_01: Wavelength Range] The spectrograph's spectral range must be wide covering in one shot the spectral range 350 -- 1750 nm (goal 320 -- 2050 nm).
\item[TLR\_02: Total Efficiency] The total efficiency at blaze peaks (including atmosphere, telescope and slit losses for a 1 arcsec seeing in V band) in the mandatory spectral bands shall be at least 25\%.
\item[TLR\_03: The Instrument] The instrument shall comprise a calibration unit, an acquisition and slit viewing unit, an optical-UV spectrograph and an infrared spectrograph. Light separation will be done by means of a dichroic.
\item[TLR\_03: Spectral Resolution and format] A spectral resolution of, at least, R=3500 must be achieved over the whole range with 1'' slit. The spectral format is either cross-dispersed Echelle, dictated by the requirements on spectral resolution and wavelength range or as alternative design, using dichroics to select ''quasi-orders'' in combination with low order gratings.
\item[TLR\_05: Spectrophotometric Accuracy] In view of its wide band--pass, the instrument shall be equipped with an ADC (Atmospheric Dispersion Compensator) for the UV-VIS spectral range for good spectrophotometric accuracy and the flexibility to orient the slit as required by the observing program
\item[TLR\_06: Calibration System] The SOXS on board calibration system shall allow performing scientific and technical calibrations needed to remove the instrument signature and to maintain the instrument.
\item[TLR\_07: Use of VLT Standard components] SOXS shall make use of VLT standard components and instrument software as far as reasonable. Should this not be the possible, well-documented or commercial products should be added.
\item[TLR\_08: Comply with NTT Interface] SOXS shall comply with the NTT telescope interface requirements and with the VLT data -- flow system.
\item[TLR\_09: On line Pipeline] The online pipeline and the data reduction package shall provide the required spectral accuracy. The online pipeline shall be able to process one typical night (series of 15--30 min exposures) in real time.
\item[TLR\_10: Safety] The instrument must be safe for people and to avoid damaging itself or other equipment, according to ESO standard.
\item[TLR\_11: Standard Observations] For standard observations, no night calibrations shall be needed to reach the performances specified.
\item[TLR\_12: Data Compatible with ESO Archive] SOXS data, raw and processed, shall be compatible with the requirements of the ESO archive.
\item[TLR\_13 Calibration Documentation] The principle applied in wavelength calibration shall be documented to allow future users of the archive to trace back the wavelength calibration.
\item[TLR\_14: Documentation] Necessary documentation should be provided by the consortium at Preliminiary Acceptance in Chile (or at a later date if mutually agreed by the parties) with the aim to allowing ESO to operate and maintain, disassemble and re--assemble the instrument for a period after the expiration of the ESO--SOXS Memorandum of Understanding. This shall also apply correspondingly to all software delivered with the instrument.
\end{description}

\section{SOXS Observing Modes}
Once on the mountain, SOXS will be offered for Service Observing, in all of its Observing Modes. Following the science case, SOXS will be a ''point and shoot'' instrument. This implies that, with the instrument still fulfilling the Science Requirements listed in Section\ \ref{sec:sciencereq}, the number of observing modes that may be selected by the user shall be as small as possible to simplify the design, test, commissioning, science verification, operation, calibration, pipeline data reduction and data quality control of SOXS. All of these tasks will benefit from having a simple instrument with few modes; also the reliability will be enhanced.
The Observing Modes that are foreseen are: {\it i}) Standard Slit Spectroscopy, with parallel use of the 2 arms; {\it ii}) Light Imaging, with Acquisition Camera.
These modes shall correspond to a single instrument setting that is fully defined just by the mode name, with only a short list of selectable slit widths. The detectors are used in staring mode with a standard setting for read speed and binning; the resulting data format is fixed.
The hardware design shall also be such that the number of user-selectable options that are left open by the design is minimized, in a way that is compatible with the main scientific goal of the instrument.
The following sub-modes shall be implemented up to the commissioning at the telescope, with the decision on the modalities with which they are offered taken by ESO: {\it a)} CCD binning factors; {\it b)} CCD detector read out speeds; {\it c)} CCD read-out window; {\it d)} De-activate one or two of the spectrographs.

\section{The instrument}
\label{sec:instrument}
\begin{figure} [ht]
   \begin{center}
   \begin{tabular}{c} 
   \includegraphics[scale=0.25]{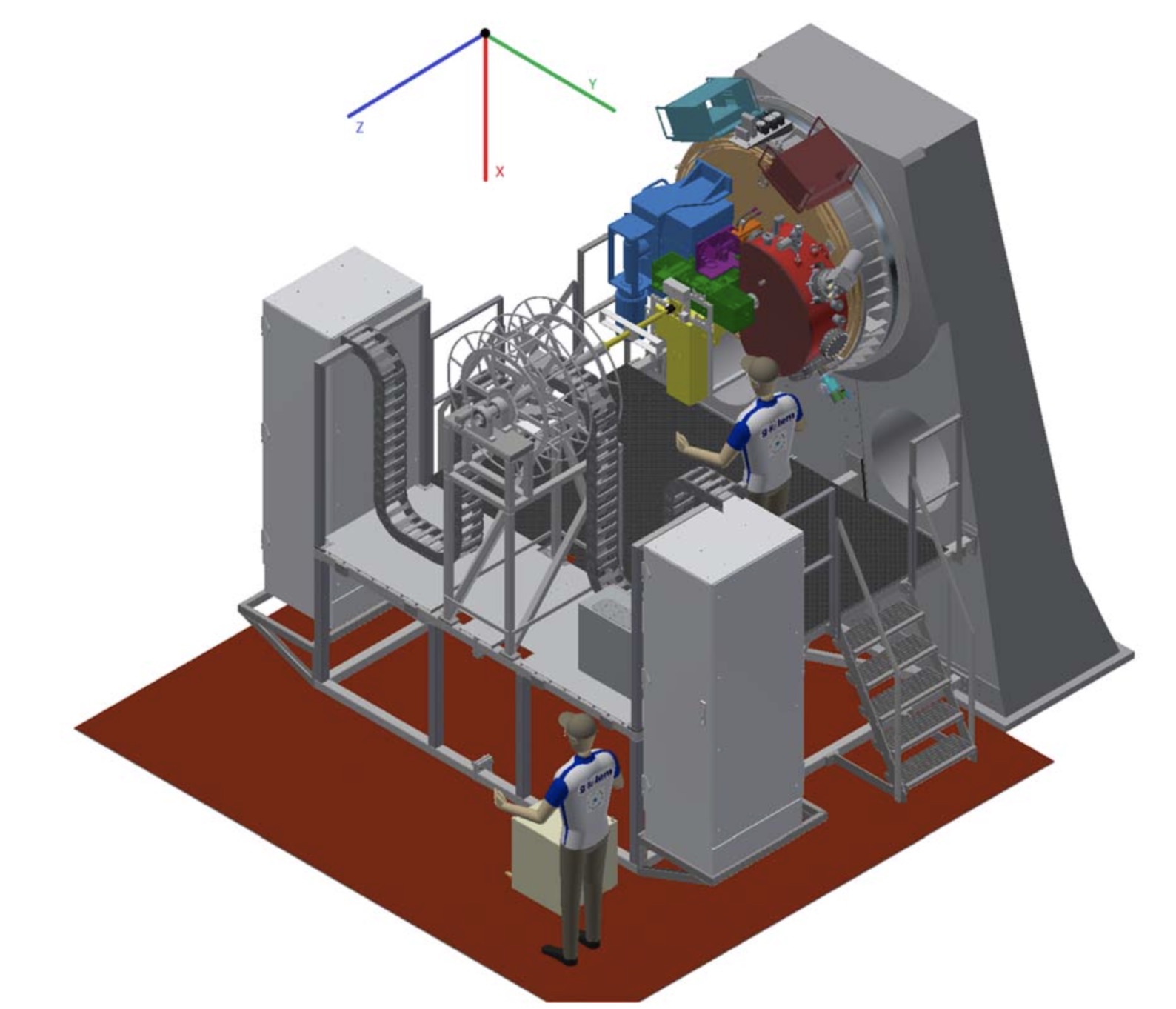}
   \end{tabular}
   \end{center}
   \caption[example] 
   { \label{fig:soxswhole} 
General SOXS view with platform and the derotator. The reference system shown is the global one.}
   \end{figure}

SOXS \cite{schipanietal2018} will simultaneously cover the electromagnetic spectrum from 0.35 to 2.0\ $\mu$m using two arms (UV--VIS and NIR) with a product slit--resolution of $\sim 4500$. The throughput should enable to reach a S/N$\sim 10$ in a 1--hour exposure of an R=20 mag point source. SOXS will be mounted at the Nasmyth focus of NTT replacing SOFI. The whole system (see Figure\ \ref{fig:soxswhole}) has three scientific arms: the UV--VIS spectrograph \cite{rubinetal2018, cosentinoetal2018}, the NIR Spectrograph \cite{vitalietal2018} and the acquisition camera \cite{brucalassietal2018}. The three main arms, the calibration box and the NTT are connected together \cite{biondietal2018} by the Common Path (CP). 

\subsection{The Common Path}
\label{sec:commpath}
\begin{figure} [ht]
   \begin{center}
   \begin{tabular}{c} 
   \includegraphics[scale=0.25]{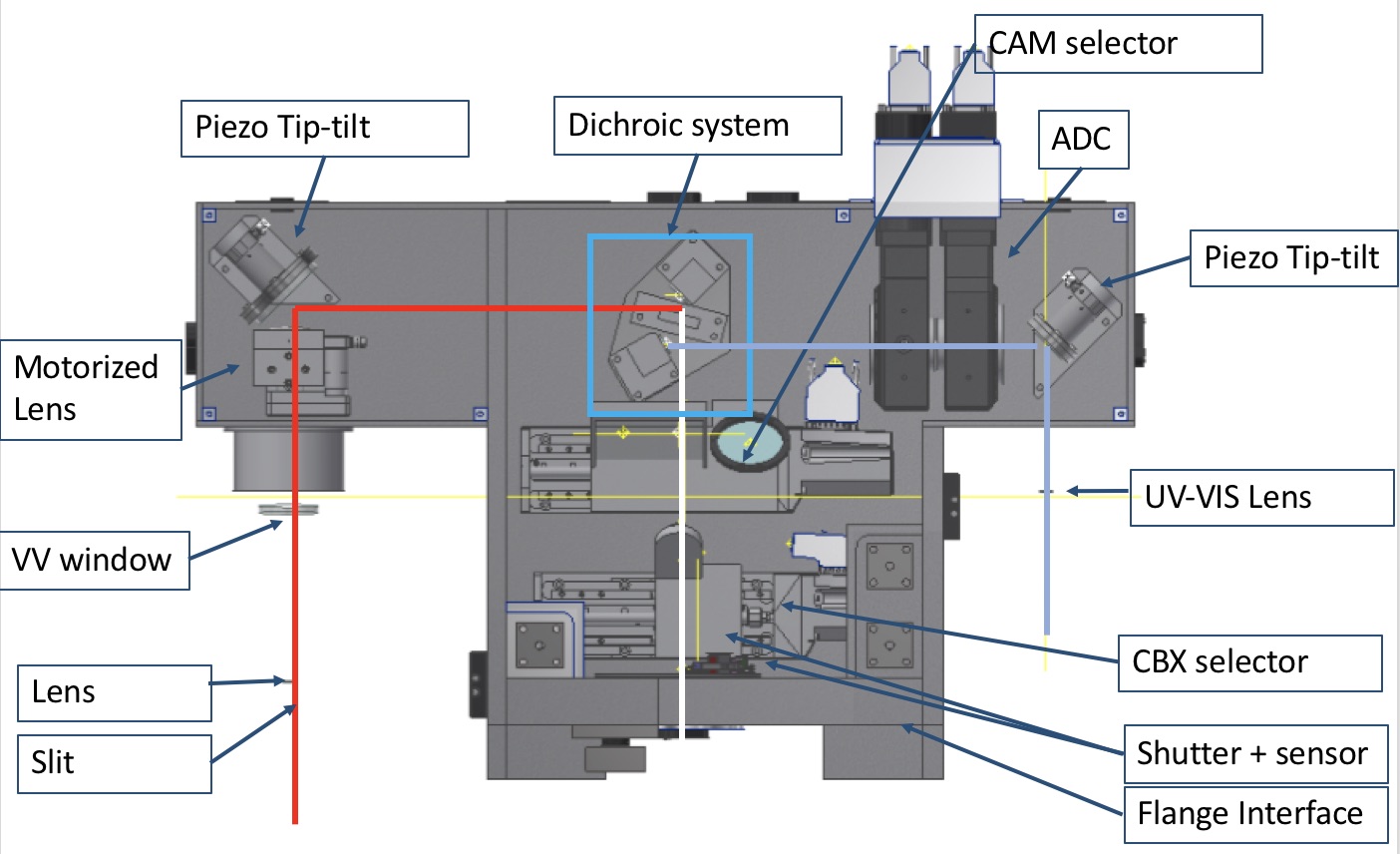}
   \end{tabular}
   \end{center}
   \caption[example] 
   { \label{fig:soxscp} 
The sketch of the Common Path sub system. In the picture the several components of the sub system are highlighted.}
   \end{figure}

The Common Path \cite{claudietal2018} relays the light from the NTT Focal Plane to the entrance of the two spectrographs (UV--VIS and NIR). It selects the wavelength range for the spectrographs using a dichroic and changes the focal ratio of the beam coming from the telescope (F/11) to one suitable for both the spectrographs. A sketch of the common path opto--mechanical design is shown in in Figure\ \ref{fig:soxscp} (overall dimensions are about $650 \times 350$\ mm) while the main CP parameters are described in Table\ \ref{tab:cpch}. 

After the dichroic, two flat folding mirrors direct the light towards two distinct UV--VIS and NIR arms. In the UV--VIS path, the light coming from the folding mirror goes to the ADC assembly, that, strictly speaking, is not composed only of two double prisms correcting the atmospheric dispersion but also of two doublets glued on the prisms. The two doublets (the first one having an aspherical surface) create a collimated beam for the ADC and transform the telescope F/11 beam into an F/6.5. After the ADC, the beam is reflected by the tip/tilt mirror mounted on a piezo stage. Finally, a field lens matches the exit pupil onto the UV--VIS spectrograph pupil. The adopted glasses assure a good transmission ($>80\ \%$) in the UV--VIS side of the spectrum. For the same reason, the dichroic is used in reflection for this wavelength range, in order to give the largest choice of materials for the substrate.

The near infrared path is very similar. It does not include an ADC as mentioned previously. A doublet reduces the telescope F/11 beam to an F/6.5 beam. The doublet has an aspherical surface deemed feasible for manufacturers. A flat tip-tilt folding mirror based on a piezo stage relays the light towards the slit. A flat window is used at the entrance of the spectrograph dewar, with a cold stop after the window itself to reduce the noise. A field lens, placed near the slit, remaps the telescope pupil on the grating of the spectrograph, as in the UV--VIS arm.

\begin{table}[ht]
\caption{Main characteristics of the SOXS common path.} 
\label{tab:cpch}
\begin{center}       
\begin{tabular}{|l|c|} 
\hline
\rule[-1ex]{0pt}{3.5ex}        Input F/\#& 11  \\
\hline
\rule[-1ex]{0pt}{3.5ex}  Field of View& $12\times12$ arcsec   \\
\hline
\rule[-1ex]{0pt}{3.5ex}  Output F/\#& 6.5  \\
\hline
\rule[-1ex]{0pt}{3.5ex}  Image Scale & $110\ \mu$m/arcsec  \\
\hline
\rule[-1ex]{0pt}{3.5ex}  Wavelength range & 350--850\ nm\ (UV--VIS); 800--2000\ nm\ (NIR)  \\
\hline 
\end{tabular}
\end{center}
\end{table}

\subsection{The NIR Spectrograph}
\label{sec:NIRSpec}
\begin{figure} [ht]
   \begin{center}
   \begin{tabular}{c} 
   \includegraphics[scale=0.15]{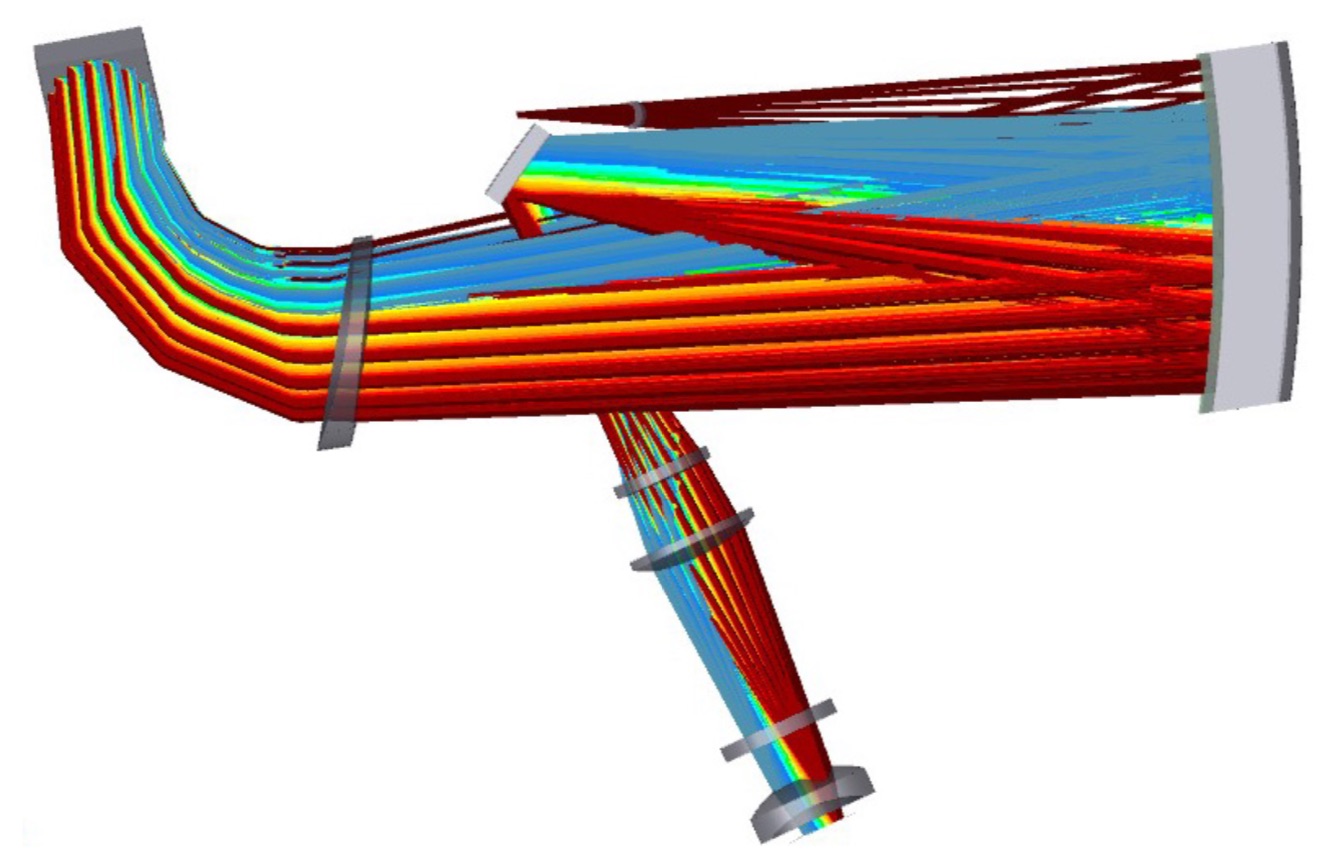}
   \includegraphics[scale=0.15]{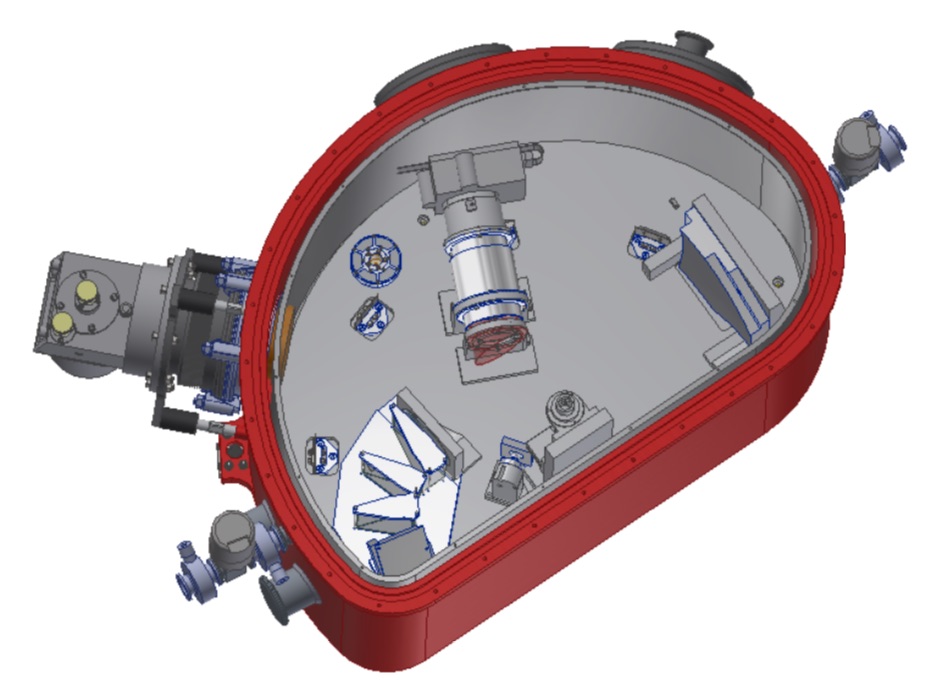}   
   \end{tabular}
   \end{center}
   \caption[example] 
   { \label{fig:soxsnir} 
The sketch of the Near Infrared Spectrograph sub system. {\it Left}: the optical layout of the spectrograph. {\it Right}: The optomechanics of the instrument.}
   \end{figure}

The near infrared spectrograph is a cross-dispersed echelle, with R=5000 (for 1 arcsec slit), covering the wavelength range from 800 to 2000\ nm with 15 orders. It is based on the 4C (Collimator Compensate Camera Chromatism) concept, characterized by a very compact layout, reduced weight of optics and mechanics, good stiffness and high efficiency. The spectrograph is composed of a double pass collimator and a refractive camera, a main grating disperser and a prism-based cross disperser. The Optical scheme and the optomechanics are shown in Figure\ \ref{fig:soxsnir} while the main parameters are summarized in Table \ref{tab:spectrographs}.

\begin{figure} [ht]
   \begin{center}
   \begin{tabular}{c} 
   \includegraphics[scale=0.25]{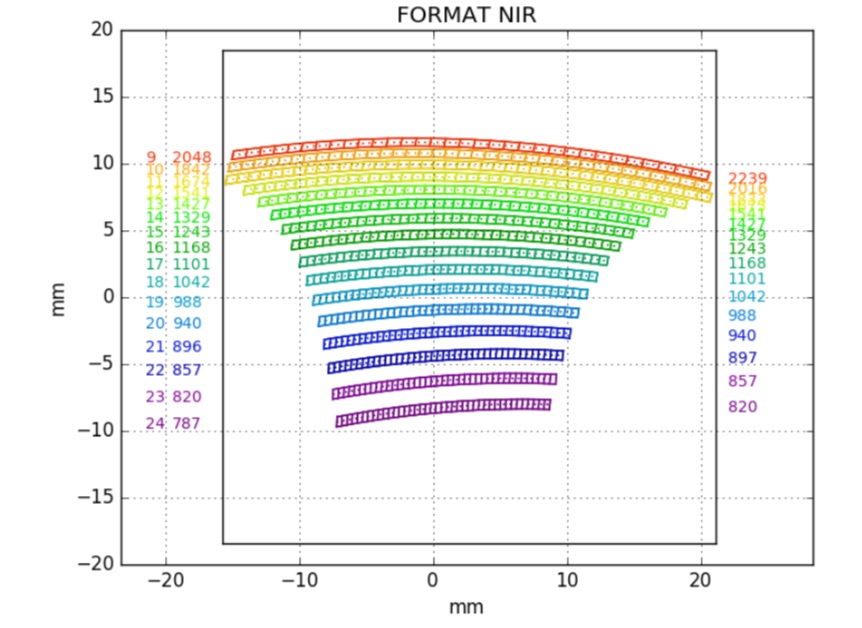}   
   \end{tabular}
   \end{center}
   \caption[example] 
   { \label{fig:soxsnirFormat} 
The spectral format  the Near Infrared Spectrograph sub system.}
   \end{figure}

During the Final Design Phase the design evolved extending the wavelength range up to $2\ \mu$m, by adding an additional order w.r.t. the previous baseline. This decision caused few changes to the overall design, e.g. the working temperature of the spectrograph optomechanics was reduced from 180\ K to 150\ K in order to maintain adequate safety margins for the thermal radiation background, supposed to be well below the dark current of the array. The design includes an accurate baffling system as well as a (removable) thermal filter at $2\ \mu$m to cut off the longer wavelength radiation. The spectral format looks like a classical cross-dispersed echelle spectrum (Figure\ \ref{fig:soxsnirFormat}).
The detector is a Teledyne H2RG array with CdZnTe substrate, removed to minimize fringing and optimize QE (Table\ \ref{tab:hawaii}) operated at 40K, i.e. below its nominal working temperature of 77K, in order to take safe margins by design to avoid possible persistency problems. The vacuum and cryogenics system is based on a Closed Cycle Cryocooler, providing a more than adequate cooling power for our relatively small spectrograph volume.

   \begin{table}[ht]
\caption{Main characteristics of the SOXS NIR Teledyne H2RG (HgCdTe) Detector.} 
\label{tab:hawaii}
\begin{center}       
\begin{tabular}{|l|c|} 
\hline
\rule[-1ex]{0pt}{3.5ex}  Format& $2048 \times 2048$  \\
\hline
\rule[-1ex]{0pt}{3.5ex}  Pixel Size& $18.0\ \mu$m \\
\hline 
\rule[-1ex]{0pt}{3.5ex}  Number of outputs & 32  \\
\hline 
\rule[-1ex]{0pt}{3.5ex}  Frame rate& 0.76 Hz with 32 outputs \\
\hline 
\rule[-1ex]{0pt}{3.5ex}  Readout noise Double Correlated& $<20$ e$^-$ rms  \\
\hline 
\rule[-1ex]{0pt}{3.5ex}  Readout noise 16 Fowler pairs& $<7$ e$^-$ rms  \\
\hline 
\rule[-1ex]{0pt}{3.5ex}  Storage Capacity @ 0.5 V& 80 Ke$^-$ \\ 
\hline 
\rule[-1ex]{0pt}{3.5ex}  Dark Current @ 77\ K& $<0.1$\ e$^-$/s \\
\hline 
\end{tabular}
\end{center}
\end{table}

\subsection{The UV--VIS Spectrograph}
\label{sec:uvvisspec}
\begin{figure} [ht]
   \begin{center}
   \begin{tabular}{c} 
   \includegraphics[scale=0.25]{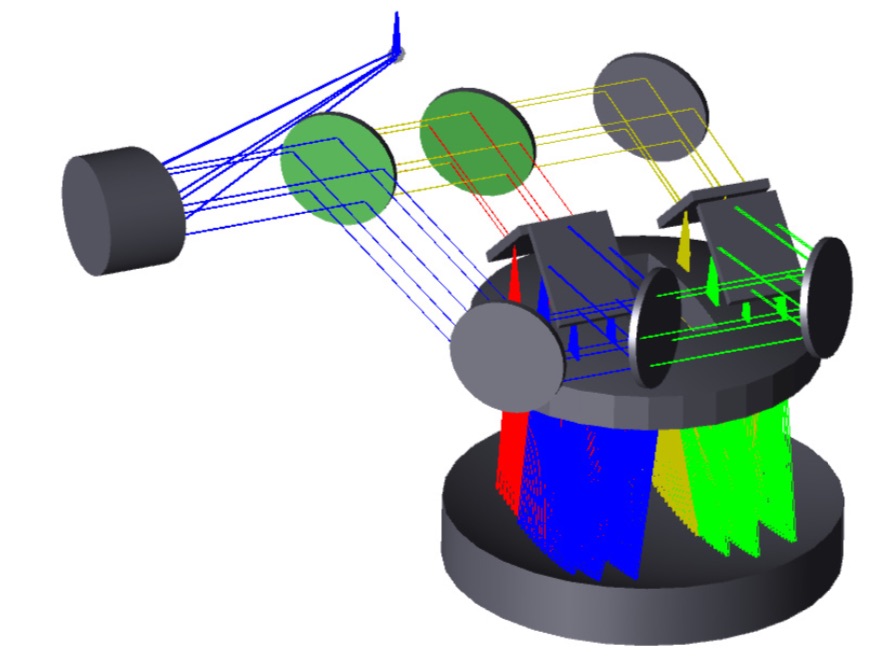}   
   \includegraphics[scale=0.20]{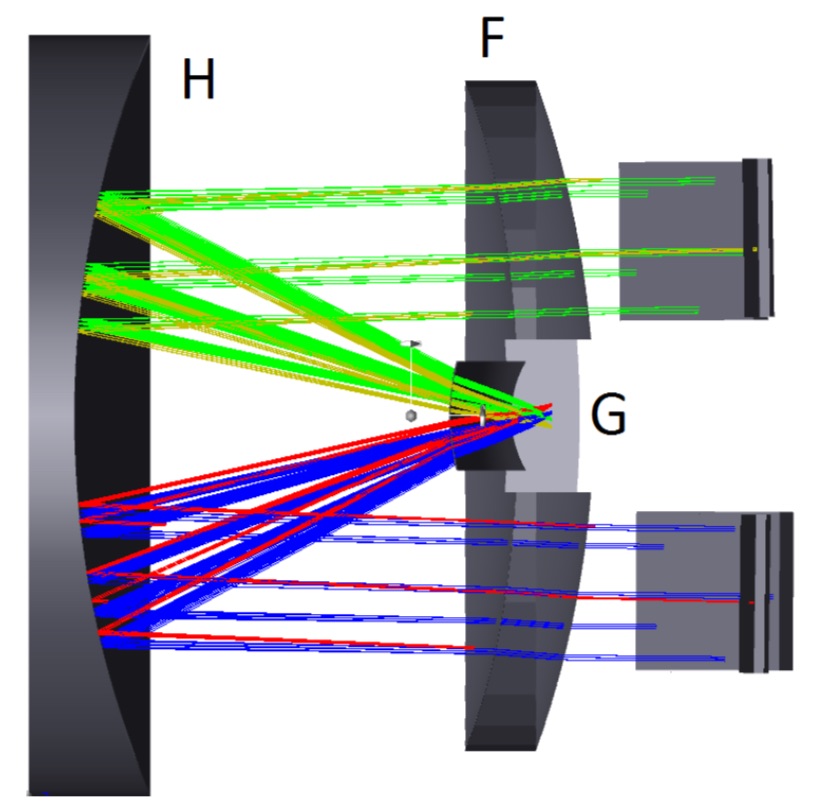}   
   \end{tabular}
   \end{center}
   \caption[example] 
   { \label{fig:soxsuvis} 
The sketch of the UV--VIS Spectrograph sub system. {\it Left}: Optical layout of the UV--VIS Spectrograph. The beam is partitioned by dichroic mirrors and imaged by a single camera. {\it Right}: Layout of camera. H, Fused silica aspheric mirrror. F, CaF$_2$ aspheric corrector and Fused silica field flattener. G, Focal plane and Detector position.}
   \end{figure}

The SOXS UV--VIS spectrograph \cite{rubinetal2018} is based on a novel concept in which the incoming beam is partitioned into four polychromatic beams using dichroic surfaces, each covering a waveband range of $\sim 100$\ nm (see Figure\ \ref{fig:soxsuvis}). Each quasi--order is diffracted by an ion--etched grating. The four beams enter a three-element catadioptric camera that images them onto a common detector. The goal of the partitioning is to maximize the overall system throughput. The optical parameters of the UV--VIS spectrograph are listed in Table\ \ref{tab:spectrographs}.

   \begin{table}[ht]
\caption{Main characteristics of the SOXS UV-VIS e2v CCD44--82 Thin Back illuminated CCD.} 
\label{tab:ccd}
\begin{center}       
\begin{tabular}{|l|c|} 
\hline
\rule[-1ex]{0pt}{3.5ex}  Format& $2048 \times 4096$  \\
\hline
\rule[-1ex]{0pt}{3.5ex}  Pixel Size& $15.0\ \mu$m \\
\hline 
\rule[-1ex]{0pt}{3.5ex}  QE at 500 nm & 90 \%  \\
\hline 
\rule[-1ex]{0pt}{3.5ex}  Charge Transfer Efficiency& 99.9995 \% \\
\hline 
\rule[-1ex]{0pt}{3.5ex}  Gain (e$^-$/ADU)& $0.6 \pm 0.1$\\
\hline 
\rule[-1ex]{0pt}{3.5ex}  RON& $<3$ e$^-$ rms  \\
\hline 
\end{tabular}
\end{center}
\end{table}

\begin{figure} [ht]
   \begin{center}
   \begin{tabular}{c} 
   \includegraphics[scale=0.25]{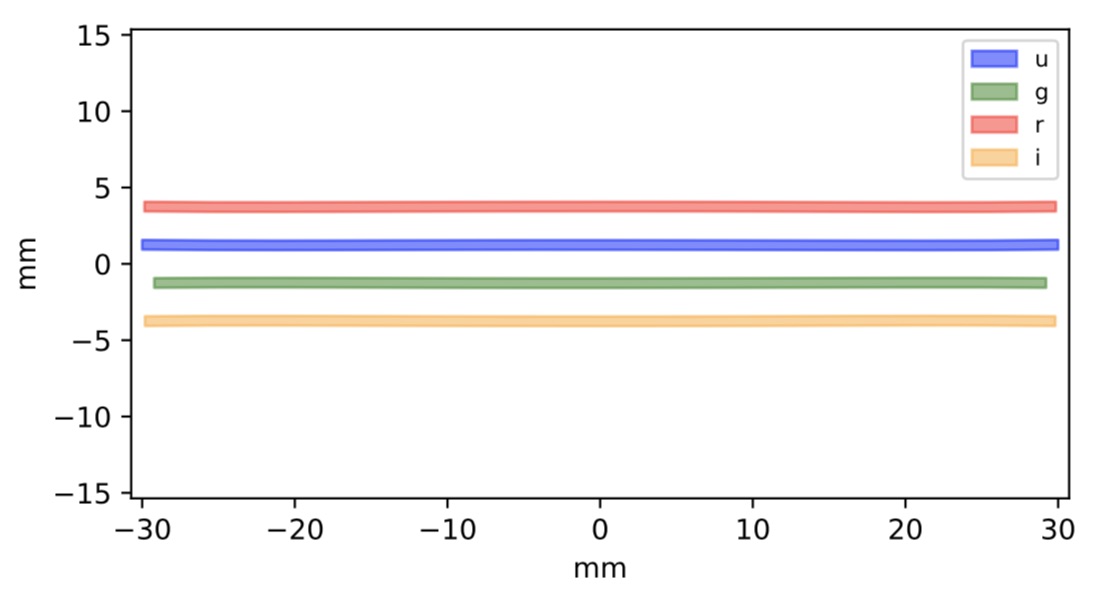}   
   \end{tabular}
   \end{center}
   \caption[example] 
   { \label{fig:soxsuvisform} Footprint on the detector including overlap regions. The different colors represent the different quasi orders. The long axis is the spectral direction.}
   \end{figure}

The detector is a Teledyne e2V CCD44-82 2k$\times$4k back illuminated CCD with $15\ \mu$m $\times\ 15\ \mu$m pixels. The main characteristics of the CCD Detector of UV--VIS Spectrograph are listed in Table \ref{tab:ccd}. The spectrum is imaged on the detector in four quasi--orders (see Figure \ref{fig:soxsuvisform}). 

Note that due to the $41^\circ$ angle of incidence of the gratings, the spatial axis is slightly rotated relative to the spectral axis. This is due to the need to orient the dispersed beam such that the central wavelength enters the corrector parallel to the optical axis of the camera. The rotation is $\sim 8^\circ$ which completes the refracted angle from $2 \times 41^\circ = 82^\circ$ to $90^\circ$. This rotation is small, is typical also of \'echelle spectra, and can be easily handled by the data reduction pipeline.

 \begin{table}[ht]
\caption{Main parameters of the two SOXS's Spectrograph.} 
\label{tab:spectrographs}
\begin{center}       
\begin{tabular}{|l|c|c|} 
\hline 
\rule[-1ex]{0pt}{3.5ex}.                       & {\bf UV--VIS }    & {\bf NIR}   \\
\hline
\rule[-1ex]{0pt}{3.5ex}  Collimator F/\#& 6.5               & 6.5     \\
\hline
\rule[-1ex]{0pt}{3.5ex}  Collimator Beam Diameter& 45\ mm  & 50\ mm \\
\hline 
\rule[-1ex]{0pt}{3.5ex}  Spectral Range& 350 -- 850\ nm       &800 -- 2000\ nm\\
\hline 
\rule[-1ex]{0pt}{3.5ex}  RS                    & 3500 -- 7000         &. 5000    \\
\hline 
\rule[-1ex]{0pt}{3.5ex}  Slit Scale           & 110 $\mu$m/arcsec    &110 $\mu$m/arcsec \\
\hline 
\rule[-1ex]{0pt}{3.5ex}  Slit Width         & 0.5 -- 1 -- 1.5 -- 5 arcsec& 0.5 -- 1 -- 1.5 -- 5 arcsec\\
\hline 
\rule[-1ex]{0pt}{3.5ex}  Slit Height         & 12 arcsec  &   12 arcsec  \\
\hline 
\rule[-1ex]{0pt}{3.5ex}  Camera Output F/\#& 3.11     & 3.7   \\
\hline 
\rule[-1ex]{0pt}{3.5ex}  Main Disperser      & 4 Custom ion etched gratings& Grating $44^\circ$, $4^\circ$ Off--Plane\\
\hline 
\rule[-1ex]{0pt}{3.5ex}  Cross Disperser       & N/A   & 3 Cleartran Prisms, apex angle $20^\circ$, \\
\rule[-1ex]{0pt}{3.5ex}       &    & $20^\circ$, $26^\circ$\\
\hline 
\rule[-1ex]{0pt}{3.5ex}  Detector Scale.        &   0.28 arcsec/pixel & 0.25 arcsec/pixel\\
\hline 
\rule[-1ex]{0pt}{3.5ex}  Working Temperature& Ambient (-5C, +20C) &  150K (40K for the last two elements\\
\rule[-1ex]{0pt}{3.5ex}   & &  of the camera and detector) \\
\hline 
\end{tabular}
\end{center}
\end{table}

 \begin{figure} [ht]
   \begin{center}
   \includegraphics[scale=0.25]{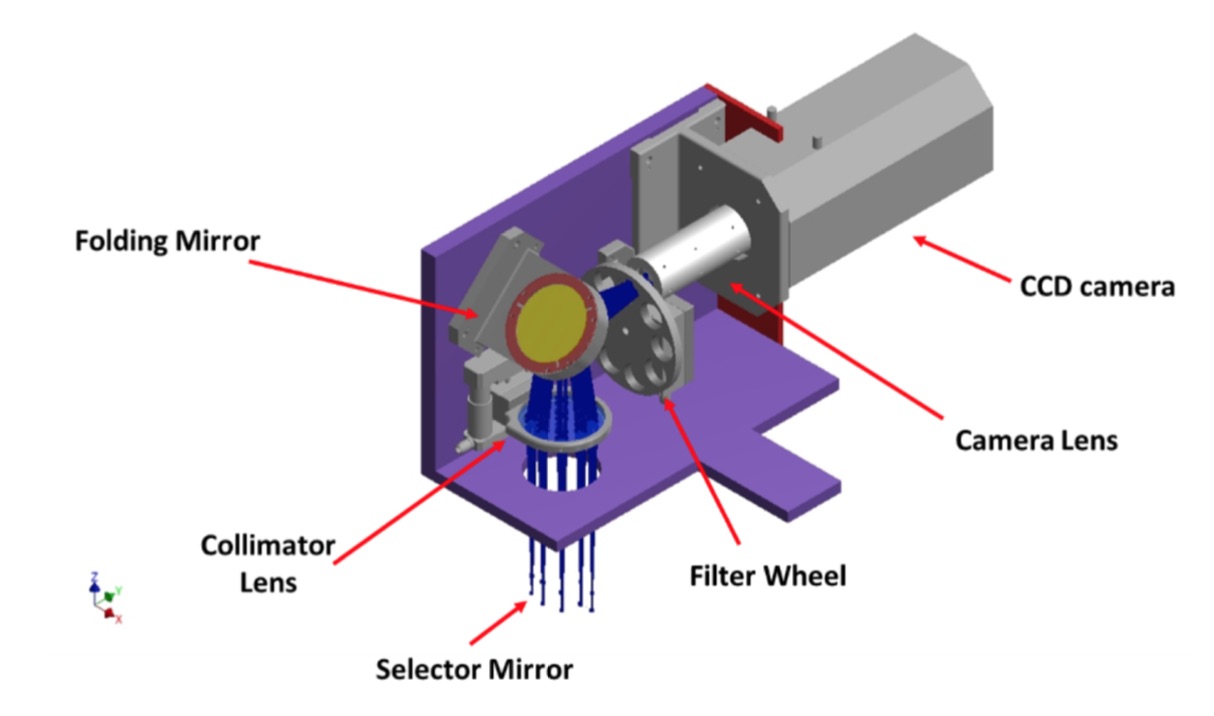}   
   \end{center}
   \caption[example] 
   { \label{fig:soxsac} Overview of the Acquisition Camera System: it consists of a collimator lens, folding mirror, filter wheel, focal reducer optics and CCD camera. All the elements are included in a structure made of 6061--T6 Aluminum.}
   \end{figure} 

\subsection{The Acquisition Camera}
\label{sec:acqucam}
The Acquisition Camera (CAM) system (see Figure\ \ref{fig:soxsac}) consists of a collimator lens, folding mirror, filter wheel, focal reducer optics and CCD camera. All the system is included in a structure made of 6061--T6 Aluminum. The CAM will receive an F/11 bean redirected from the telescope focal plane through the so called CAM selector placed in the CP. The CAM selector (see Figure\ \ref{fig:soxscp}) is based on a stage that at the level of the Nasmyth focal plane, carries a single mirror with three positions for different functions and a pellicle beam-splitter. The mirror and the pellicle beam--splitter are tilted at $45^\circ$ and they direct light from sky (the mirror) or from the slits (the pellicle) to the CAM optics. The focal plane is placed at 500mm from the NTT Nasmyth interface with a plate scale of 5.359 arcsec/mm.
The main functionalities of the CAM system are described below, along with the corresponding CAM selector configurations: {\it a)} CP pellicle beamsplitter: the semitransparent pellicle beamsplitter, inclined towards the instrument, allows to use the CAM system as a slit viewing camera (with calibration lamp on). Selecting different filters, one is able to use the CAM system as a selective on-slit viewer. During daytime visually co-alignment of the slits could be possible by illuminating them with the calibration lamp through the pellicle; {\it b)} CP flat $45^\circ$ mirror with 3 selectable positions: {\it Acquisition and Imaging}: this position allows for a light imager mode of the instrument with a complete set of broad band filters: 5 SDSS filters (u, g, r, i, z), Y to reproduce the LSST filter set, V (Johnson) and a free position for the slit viewer. {\it Artificial star}: this configuration has been implemented for maintenance reasons and will be used in daylight time.  {\it Monitoring (Spectroscopy)}: this mode will be selected during the science exposure. The $45^\circ$ mirror is translated to place a hole on the optical axis. This passes an unvignetted field of 15 arcsec (corresponding to 2.805mm) to the spectrograph slits. On the CAM it is possible to recover the peripheral field, but with a hole in the center. Off--axis secondary guiding on a peripheral object in the periphery will also be implemented during science exposures. 

The detector is an Andor iKon-M 934 Series Camera with a CCD sensor model BEX2-DD ($1024 \times 1024$) that assures a broadband coverage and a high NIR Quantum Efficency. The pixel readout rate ranges among 5, 3, 1, 0.05 MHz with a frame rate of 4.4 fps (full frame). The characteristic of the detector are: Read-out noise= 2.9e$^-$ and a dark current of 0.00012 e$^-$/pixel/sec at $-100.0^\circ$ C.

 \begin{figure} [ht]
   \begin{center}
   \includegraphics[scale=0.20]{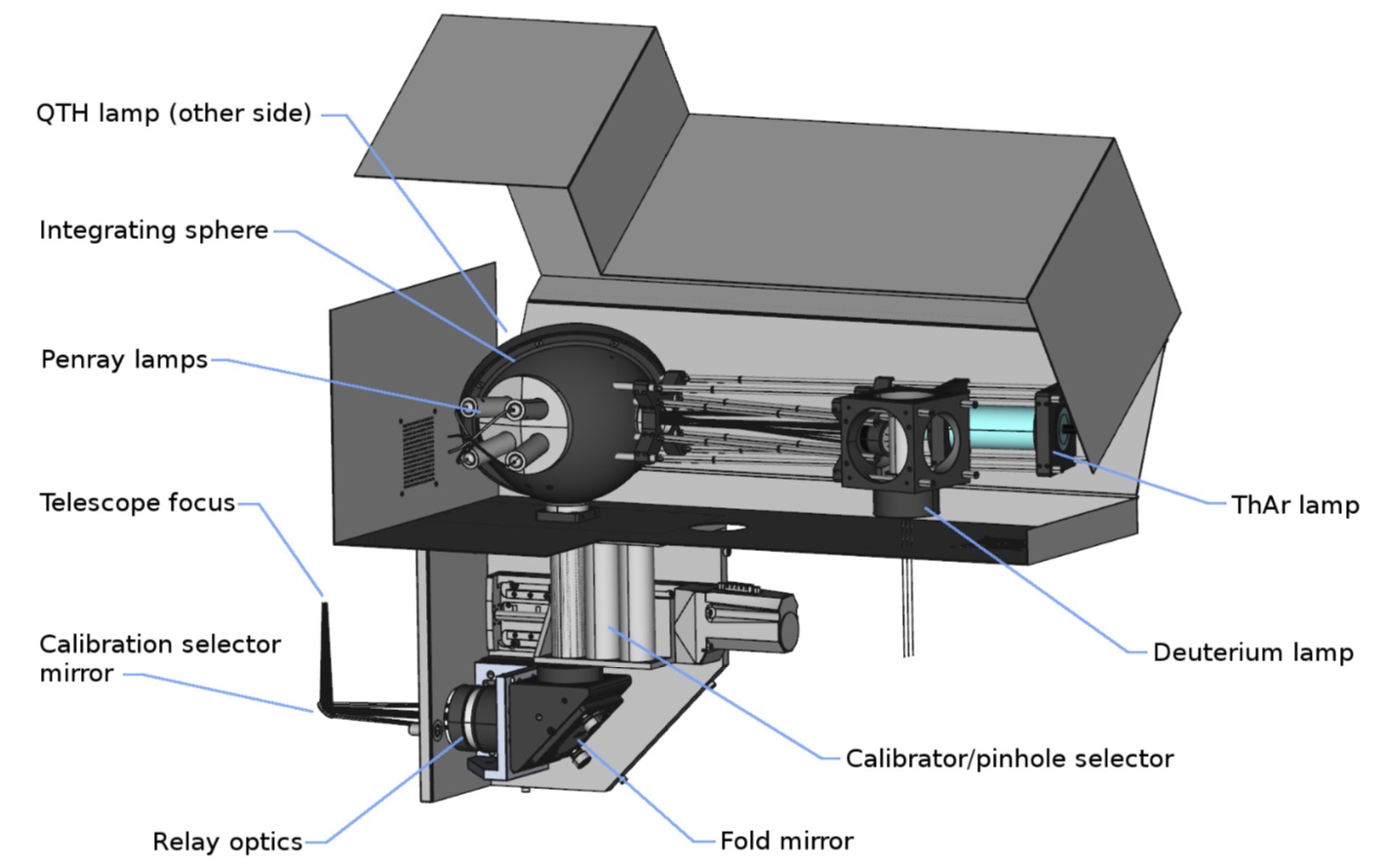}   
   \end{center}
   \caption[example] 
   { \label{fig:soxscalib} The calibration unit.}
   \end{figure} 

\subsection{The calibration Box}
\label{sec:calbox}
The calibration unit (Figure \ref{fig:soxscalib}) is designed to provide the calibration spectra to remove the instrument signatures and convert the observed spectrum into one with physical units (in wavelength and flux). The calibration spectra are generated using a synthetic light source, adopting an integrating sphere equipped with lamps suitable for wavelength and flux calibrations across the full wavelength range of the instrument (350-2000 nm). The following lamps are used: quartz-tungsten-halogen (QTH) lamp, for flux calibration 500-2000 nm;  deuterium (D2) lamp, for flux calibration 350-500 nm (used simultaneously with QTH lamp for UV-VIS arm
flux calibration); Ne Ar Hg Xe pen ray lamps bundled together, for NIR wavelength calibration. The individual lamps are controlled to operate together as one lamp. ThAr hollow cathode lamp, for UV-VIS wavelength calibration. At the exit port of the sphere, the light is diffused in order to achieve uniform illumination at the slit plane. Relay optics are foreseen between the light source and the focal plane. In addition, a fold mirror is foreseen in the light path to accommodate the optical design to the space and mechanical constraints. This relay system reimages the exit pupil of the integrating sphere to the focal plane, maintaining uniform illumination at the slit. A stage-controlled calibration pick-up mirror is placed in the light path before the telescope focus, which enables the observer to switch the spectrograph input light between telescope and calibration sources. The system is also equipped with a pinhole mask mounted on a linear motor stage, located near the exit port of the integrating sphere. This enables the creation of an artificial star for engineering purposes.

\subsection{Electronics and Cryogenics}
\label{sec:elecry}
\begin{figure} [hb]
   \begin{center}
   \includegraphics[scale=0.40]{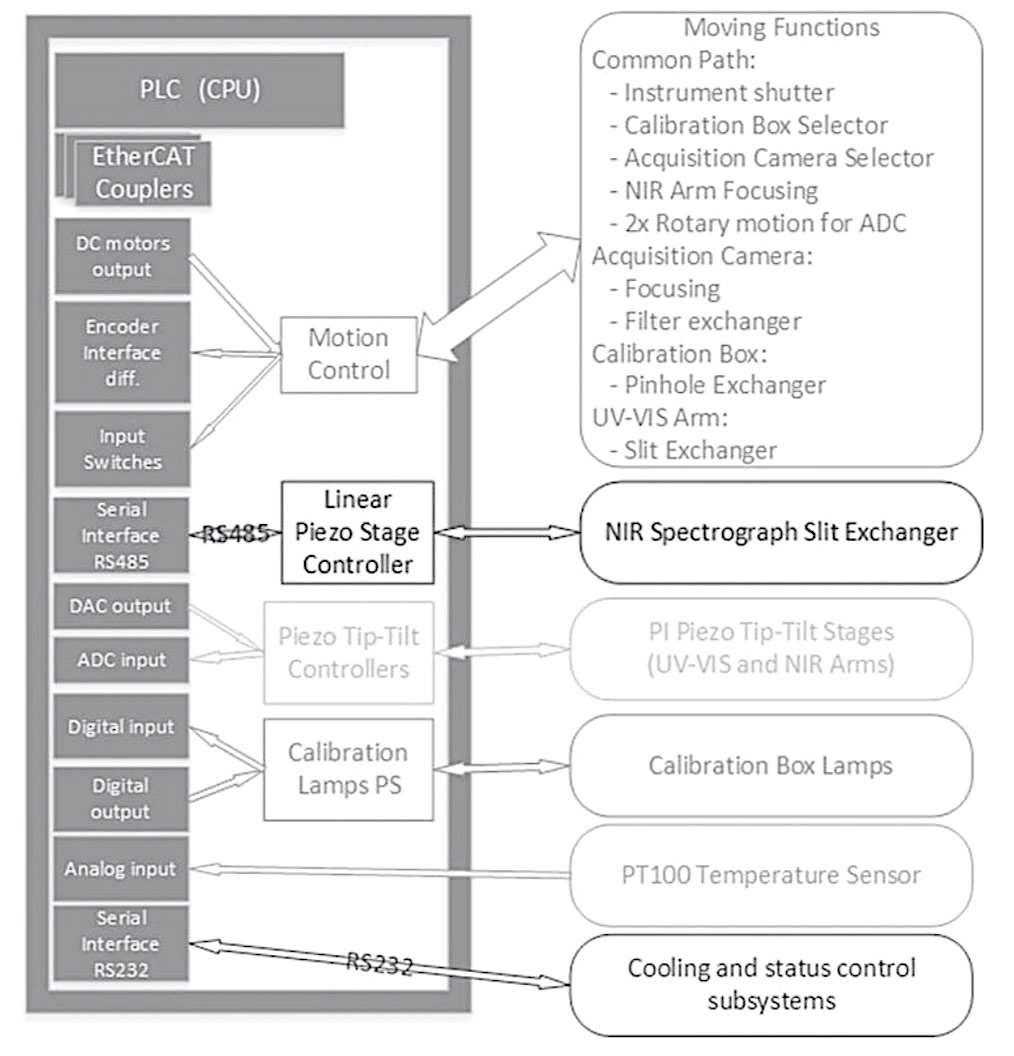}   
   \end{center}
   \caption[example] 
   { \label{fig:soxsice} Overview of the instrument control electronic of SOXS.}
   \end{figure}

The Instrument Control Electronics (ICE) \cite{capassoetal2018} is based on one main PLC and some I/O modules connected to all subsystems. The modules are connected to the main PLC via the EtherCAT fieldbus.
The PLC offers an OPC-Unified Architecture interface on the LAN. The Instrument Software (INS) \cite{riccietal2018} installed on the Instrument Workstation (IWS) uses this protocol to send commands and read the status to/from the PLC. Figure\ \ref{fig:soxsice} shows the overview of the ICE of SOXS.

\begin{figure} [ht]
   \begin{center}
   \includegraphics[scale=0.30]{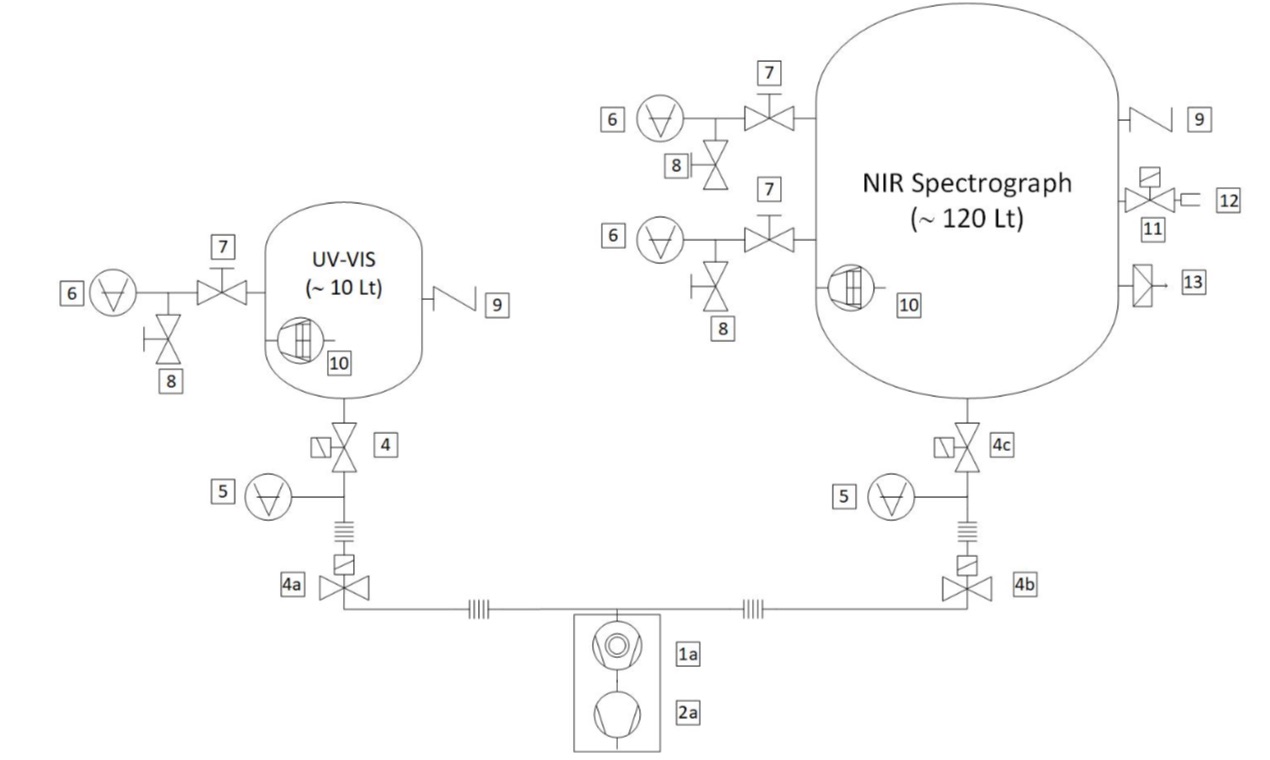}   
   \end{center}
   \caption[example] 
   { \label{fig:soxscryo} Overview of the SOXS vacuum system.}
   \end{figure}

The SOXS Vacuum system (Figure\ \ref{fig:soxscryo}) is designed with a system of pre-vacuum and turbo-molecular pumps that is common to the two arms. They will be located on the Nasmyth platform at a very short distance and connected only when vacuum operations are needed. This solution reduces the load on the instrument, leaving anyway the possibility to connect the turbo-molecular pump on board if needed. The commercial components have been selected in agreement to ESO standards. In the visible arm, the small CCD cryostat will be based on the Continuous Flow Cryostat concept, successfully adopted in several ESO projects. A Closed Cycle CryoCooler will be adopted in the near infrared instrument cryogenic system.

\subsection{Instrument Software (INS)}
\label{sec:ins}
The software \cite{riccietal2018} is based on the VLT Instrumentation Common Software. This solution led to standard management of instrument components, observation, calibration and maintenance procedures. Observation procedures for spectroscopy, imaging, calibration and maintenance are developed as templates, which are executed by the Broker of Observation Blocks (BOB) through commands sent to the OS. The work is in progress for the usual software subsystems, i.e. the Instrument Control Software, the Detector Control Software, and the Observation and Maintenance Software.

\section{The Assembly, Integration and Test processes}
\label{sec:ait}
The instrument construction philosophy is that all sub-systems will be built in different institutes and will be moved to the integration site, at the INAF Astronomical Observatory of Padova, before the Preliminary Acceptance in Europe (PAE). Just before the shipping to Padova, all sub-systems have to be tested in order to verify SOXS science and technical requirements. This phase is an internal review named Assembly Readiness Review (ARR). After the PAE, the instrument will be dismounted and moved to Chile where it will be re-integrated and tested for acceptance. 
Eventually, the instrument will undergo to a commissioning phase and science validation phase. 
During the different phases mainly four type of tests are foreseen: 
\begin{itemize}
\item Alignment: all the procedures that allow having the throughput as high as possible for the sub-systems and the instrument. It will be necessary to check the alignment every time the sub-systems and the instrument will be partially dismounted and successively re--integrated.
\item Functional tests: all the tests devoted to verify the full functionality of the sub-systems and of the instrument. These tests will verify the electro-mechanics movements, sensors and telemetry.  
\item Verification tests: all the tests that verify, depending by the test phases, all the internal and external (including those towards the telescope) interfaces. These tests will also verify the compliance of the sub--systems and the instrument with all environmental requirements.
\item Science tests: all the tests that will verify the compliance of the sub--systems and the instrument with science and those technical requirements that concern the scientific output of the sub-systems and the instrument.
\end{itemize}
After its reintegration and the successful completion of all functional tests, the instrument will be ready for the commissioning (''First Light'') observations.
Two commissioning periods (COMM1 and COMM2) are foreseen. The two periods will be separated by a performance assessment period. The timing of all operation after the PAE is described in Table\ \ref{tab:soxstimetable}.
The gap between the two commissioning periods is needed to allow for a performance assessment of the results of the COMM1 run and to make the necessary modifications on the instrument hardware and/or software. COMM2 will start when all important actions on SOXS based on the results of COMM1 are successfully executed.

COMM1 will be used to characterize the functionality of SOXS on the sky in its complete configuration. In particular COMM1 will verify: the instrument functionality (hardware and software), to ensure that the instrument is responding to all commands and is fully functional in all modes; the instrument performance (hardware and software), to ensure that the instrument is performing as specified, verifying fundamental parameters for instrument operation. On--sky tests will be performed using templates and observation blocks; the interfaces with the NTT system (TCS, Data Flow, Observing Software) of the INS and of the reduction and analysis software.

COMM2 period shall serve to prove that the instrument complies with all requirements, in particular with those that could not be verified in laboratory but only on the sky. The instrument performances on the sky must be fully characterized and the instrument must prove to work in compliance with normal La Silla operations. COMM2 will start when all important corrections to SOXS based on the results of COMM1 have been executed. In particular COMM2 will verify: the instrument integration with the NTT system in the science operations scenario of La Silla; hardware and software functionality and complete characterization of the instrument; on--sky performances to ensure that the instrument is performing as specified.

COMM2 will be followed by the Science Verification for the early scientific evaluation of the instrument capabilities. Science verification programs to demonstrate the capabilities of SOXS to the community will be carried out by the SOXS Principal Investigator. The observed data will immediately be made public according to the Science Verification ESO policy.


   \begin{table}[ht]
\caption{Installation and commissioning (COMM) schedule} 
\label{tab:soxstimetable}
\begin{center}       
\begin{tabular}{|l|c|} 
\hline
\rule[-1ex]{0pt}{3.5ex}  Nasmyth Room Available & 2020--09--11 \\
\hline
\rule[-1ex]{0pt}{3.5ex}  SOXS @ La Silla&  2020--09--11\\
\hline 
\rule[-1ex]{0pt}{3.5ex}  Integration \& functionality Tests& 2020--09--14 to 2010--10--16 \\
\hline 
\rule[-1ex]{0pt}{3.5ex}  COMM1 @ NTT &  2020--10--19 to 2020--10--30 \\
\hline 
\rule[-1ex]{0pt}{3.5ex} Performance Assessment (Europe)& 2020--11--02 to 2021--01--01\\
\hline 
\rule[-1ex]{0pt}{3.5ex} COMM2 @ NTT& 2021--01--04 to 2021--01--15  \\
\hline 
\rule[-1ex]{0pt}{3.5ex} Science Verification& 2021--02--15 to 2021--02--26  \\
\hline 
\end{tabular}
\end{center}
\end{table}

\section{Conclusions}
\label{sec:conc}
SOXS will be the work-horse for the study of transients of the ESO La Silla observatory. The project is going to conclude the final design phase in 2018. Afterward, the plan is to start operations in 2021. After the completion of the instrument, the consortium will still be involved in the regular operation phase.


\begin{thebibliography}{99}

\bibitem{vernetetal2011} J. Vernet et al., ''X-shooter, the new wide band intermediate resolution spectrograph at the ESO Very Large Telescope'', A\&A {\bf 536}, A105 (2011).

\bibitem{dezeeuw2016}T. de Zeeuw, "Reaching new heights in astronomy -- ESO long term perspectives", The Messenger 166, 2-27 (2016).

\bibitem{mayoretal2003} Mayor M., Pepe F., Queloz D., Bouchy F., Ruprecht et al., ''Setting New Standards with HARPS'', The Messenger, {\bf 114}, 20, (2003)

\bibitem{wildietal2017}Wildi, F.; Blind, N.; Reshetov, V.; Hernandez, O.; Genolet, L.; Conod, U.; Sordet, M.; Segovilla, A.; Rasilla, J. L.; Brousseau, D. et al., ''NIRPS: an adaptive-optics assisted radial velocity spectrograph to chase exoplanets around M-stars'', Proceedings of the SPIE, {\bf 10400}, id. 1040018 15 pp. (2017)

\bibitem{smarttetal2015}S. J. Smartt, et al.,''PESSTO: survey description and products from the first data release by the Public ESO Spectroscopic Survey of Transient Object'', A\&A {\bf 579}, A40 (2015).

\bibitem{gaiacollaborationetal2016} Gaia Collaboration et al., ''the GAIA Mission'', A\&A, {\bf 595}, A1, (2016)

\bibitem{kaiseretal2010} Kaiser, N., Burgett, W., Chambers, K., et al. 2010, in Proc. SPIE, Vol. {\bf 7733}, Ground-based and Airborne Telescopes III, 77330E

\bibitem{bellm2018} Bellm E., ''Life Beyond PTF'', arXiv: 1802.10218 (2018) 

\bibitem{ivezicetal2008}Ivezic, Z., et al., arXiv:0805.2366, (2008)

\bibitem{laureijsetal2012}Laureijs, R.; Gondoin, P.; Duvet, L.; Saavedra Criado, G. et al.,''Euclid: ESA's mission to map the geometry of the dark universe'', Space Telescopes and Instrumentation: Optical, Infrared, and Millimeter Wave. Proceedings of the SPIE, {\bf 8442}, article id. 84420T, (2012)

\bibitem{gehrelsetal2004}Gehrels, N. et al., 'The Swift Gamma-Ray Burst Mission', Astrophys. J., {\bf 611}, 1105--1120 (2004)

\bibitem{atwoodetal2009}Atwood, W. B., Abdo, A. A., Ackermann, M., et al. , ApJ, {\bf 697}, 1071, (2009)

\bibitem{lorenz2004}Lorenz, E. , NewA. Rev., {\bf 48}, 339, (2004)

\bibitem{theligocollaborationetal2011} The Ligo Scientific Collaboration et al, ''A Gravitational wave observatory operating beyond the quantum shot noise limit'', Nature Physics, {\bf 7}, 962, (2011)

\bibitem{kulkarni2012} Kulkarni S. R., ''Cosmic Explosions (Optical)'' in New Horizons in Time Domain Astronomy, IAU Symposium, Griffin, E. and Hanisch, R. and Seaman, R., {\bf 285}, 55--61, (2012)

\bibitem{jonkeretal2012} Jonker, P. G.; Miller--Jones, J. C. A.; Homan, J.; Tomsick, J.; Fender, R. P.; Kaaret, P.; Markoff, S.; Gallo, E., ''The black hole candidate MAXI J1659--152 in and towards quiescence in X--ray and radio'', MNRAS, {\bf 423}, 3308, (2012)

\bibitem{mayorandqueloz1995} Mayor M., Queloz D., ''A Jupiter--mass companion to a Solar -- type Star, Nature, {\bf 378}, 355 (1995)

\bibitem{brown2001} Brown T.M., ''Transmission Spectra  as Diagnostics of Extrasolar Giant Planet Atmospheres'', Ap. J., {\bf 553}, 1006, (2001)

\bibitem{bodeandevans2008}Bode M.F., Evans A., in Classical Novae, 2nd Edition. Edited by M.F. Bode and A. Evans. Cambridge Astrophysics Series, No. 43, Cambridge: Cambridge University Press, (2008)

\bibitem{henzeetal2013}Henze et al., Classical Novae as Supersoft X-ray Sources in the Andromeda Galaxy, in: Binary Paths to Type Ia Supernovae Explosions, Proceedings of the International Astronomical Union, IAU Symposium, Volume 281, p. 105 (2013)




\bibitem{paynegaposchkin1957} Gaposchkin, C.H., Payne, The Galactic Novae, Amsterdam, North--Holland Pub. Co.; New York, Interscience Publishers, 1957.

\bibitem{livioandtruran1994} Livio M., Truran, J. W., On the interpretation and implications of nova abindances: an abundance of riches or an overabundance of enrichments, ApJ, {\bf 425}, 797 (1994)

\bibitem{abdoetal2010} Abdo A.A., Ackermann, M., Ajello M., Atwood W.B., Baldini L., Ballet J., Barbiellini G., Bastieri D., Baughman B.M., Bechtol, K. et al., Fermi Gamma -- Ray Imaging of a Radio Galaxy, Science, {\bf 329}, 817, (2010)

\bibitem{sokerandkashi2012} Soker N., Kashi A., Formation of Bipolar Planetary Nebulae by Intermediate -- luminosity Optical Transient, ApJ, {\bf 746}, 100, (2012)



\bibitem{maozetal2014} Maoz D., Mannucci F., \& Nelemans G., Observational Clues to Progenitors of Type Ia Supernovae, ARA\&A, {\bf 52}, 107, (2014)

\bibitem{hegeretal2003} Heger A., Fryer C.L., Woosley S.E., Langer N., Hartmann D.H., How Massive Single Stars End Their Life, ApJ, {\bf 591}, 288, (2003)

\bibitem{stritzingeretal2015} Stritzinger M.D., Valenti S., Hoeflich P., Baron E., Phillips M. M., Taddia F., Foley R.J., Hsiao E.Y., Jha S.W., McCully C., et al., Comprehensive observations of the bright and energetic Type Iax SN 2012Z: Interpretation as a Chandrasekhar mass white dwarf explosion, A\&A, {\bf 573}, A2, (2015)

\bibitem{fynboetal2006} Fynbo, J. P. U.; Watson, D.; Th\"one, C. C.; Sollerman, J.; Bloom, J. S.;  et al., No supernovae associated with two long-duration $\gamma$--ray bursts, Nature, {\bf 444}, 1047, (2006)

\bibitem{dellavalleetal2006} Della Valle, M.; Chincarini, G.; Panagia, N.; Tagliaferri, G.; Malesani, D.; Testa, V.; Fugazza, D.; Campana, S.; Covino, S.; Mangano, V. et al., An enigmatic long-lasting $\gamma$--ray burst not accompanied by a bright supernova, Nature, {\bf 444}, 1050, (2006)




\bibitem{lorimeretal2007}Lorimer, D. R.; Bailes, M.; McLaughlin, M. A.; Narkevic, D. J.; Crawford, F., A Bright Millisecond Radio Burst of Extragalactic Origin, Science, {\bf 318}, 777, (2007)

\bibitem{thorntornetal2013} Thornton, D.; Stappers, B.; Bailes, M.; Barsdell, B.; Bates, S. et al., A Population of Fast Radio Bursts at Cosmological Distances, Science, 341, {\bf 53}, (2013)

\bibitem{kulkarnietal2014} Kulkarni, S. R.; Ofek, E. O.; Neill, J. D.; Zheng, Z.; Juric, M., Giant Sparks at Cosmological Distances?, ApJ, {\bf 797}, 70, (2014)

\bibitem{abbotetal2017} B. P. Abbott, et al., Multi-messenger observations of a binary neutron star merger, ApJL {\bf 848}:L12, (2017).

\bibitem{pianetal2017} E. Pian, et al., "Spectroscopic identification of r-ptocess nucleosynthesis in a double neutron star merger", Nature {\bf 551}, 67-70, (2017).

\bibitem{smarttetal2017} S. Smartt, et al., "A kilonova as the electromagnetic counterpart to a gravitational-wave source", Nature {\bf 551}, 75-79, (2017).

\bibitem{piranetal2013} Piran, T.; Nakar, E.; Rosswog, S.,  The electromagnetic signals of compact binary mergers, MNRAS, {\bf 430}, 2121, (2013)

\bibitem{barnesandkasen2013} Barnes, J.; Kasen, D., Effect of a High Opacity on the Light Curves of Radioactively Powered Transients from Compact Object Mergers,  ApJ, {\bf 775}, 18, (2013)

\bibitem{singeretal2013} Singer, L. P.; Cenko, S. B.; Kasliwal, M. M.; Perley, D. A.; Ofek, E. O. et al., Discovery and Redshift of an Optical Afterglow in 71 deg$^2$: iPTF13bxl and GRB 130702A, ApJ, {\bf 776}, L34, (2013)

\bibitem{singeretal2015} Singer, L. P.; Kasliwal, M. M.; Cenko, S. B.; Perley, D. A. The Needle in the 100 deg$^2$ Haystack: Uncovering Afterglows of Fermi GRBs with the Palomar Transient Factory, ApJ, {\bf 806}, 52, (2015)

\bibitem{schipanietal2018} P. Schipani, S. Campana, R. Claudi, H. U. K\"aufl, M. Accardo, M. Aliverti,  et al., SOXS: a wide band spectrograph to follow up transients, Proc. SPIE 10702, 107020F, doi: 10.1117/12.2307349, (2018).

\bibitem{rubinetal2018} Rubin A., Ben--Ami S., Hershko O., Rappaport M., Diner O., Gal--Yam A., Campana S., Claudi R., Schipani P., Aliverti M., et al., MTIS: the Multi--Imaging Transient Spectrograph for SOXS, Proc. SPIE 10702, 107022Z, doi: 10.1117/12.2313338, (2018).

\bibitem{cosentinoetal2018} Cosentino R., Aliverti M., Scuderi S., Campana S., Schipani P., Baruffolo A., Ben--Ami S., Mehrgan L.H., Ives D., et al., The VIS detector system of SOXS, Proc. SPIE 10702, 107022J, doi: 10.1117/12.2312539,  (2018).

\bibitem{vitalietal2018} Vitali F., Aliverti M., Capasso G., D'Alessio F., Munari M., Riva M., Scuderi S., Zanmar Sanchez R., Campana S., Schipani P., et al., The NIR Spectrograph for thr new SOXS Instrument at the NTT, Proc. SPIE 10702, 1070228, doi: 10.1117/12.2311466, (2018).

\bibitem{brucalassietal2018} Brucalassi A., Araiza -- Duran A., Pignata G., Campana S., Claudi R., Schipani P., Aliverti M., Baruffolo A., et al., The Acquisition Camera System for SOXS at NTT, Proc. SPIE 10702, 107022M, doi: 10.1117/12.2312677, (2018).

\bibitem{biondietal2018} Biondi F., Claudi R., Marafatto L., Farinato J., Magrin D., Ragazzoni R., Campana S., Schipani P., Aliverti M., Baruffolo A., et al., The Assembly, Integration and test activities for the new SOXS instrument at NTT, Proc. SPIE 10702, 107023D, doi: 10.1117/12.2313944, (2018).

\bibitem{claudietal2018} R. Claudi, M. Aliverti, F. Biondi, M. Munari, R. Z\'anmar S\'anchez, et al., The Common Path of SOXS (Son of X-Shooter), Proc. SPIE 10702, 107023T, doi: 10.1117/12.2314576, (2018).

\bibitem{capassoetal2018} Capasso, G.; Colapietro, M.; D'Orsi, S.; Schipani, P.; Aliverti, M.; Kuncarayakti, H.; Scuderi, S.; Coretti, I.; Campana, S.; Claudi, R. et al., SOXS Control Electronics Design, Proc. SPIE 10707, 107072H, doi: 10.1117/12.2312780, (2018).

\bibitem{riccietal2018} Ricci, D.; Baruffolo, A.; Salasnich, B.; Fantinel, D.; Urrutia, J.; Campana, S.; Claudi, R.; Schipani, P. et al., Architecture of the SOXS instrument control software, Proc. SPIE 10707, 107071G, doi: 10.1117/12.2310092, (2018).

\end{thebibliography}
\end{document}